\documentclass[preprint,amsmath,amssymb,superscriptaddress,unsortedaddress]{revtex4}

\usepackage{epsf}
\usepackage{graphicx}
\usepackage{color}
\begin{document}

\title{Observation of quasiparticles for phase incoherent $d$-wave pairing in Bi2212}

\author{Takeshi Kondo}
\affiliation{ISSP, University of Tokyo, Kashiwa, Chiba 277-8581, Japan}

\author{W.~Malaeb} 
\affiliation{ISSP, University of Tokyo, Kashiwa, Chiba 277-8581, Japan}

\author{Y.~Ishida} 
\affiliation{ISSP, University of Tokyo, Kashiwa, Chiba 277-8581, Japan}

\author{T.~Sasagawa} 
\affiliation{Materials and Structures Laboratory, Tokyo Institute of Technology, Yokohama, Kanagawa 226-8503, Japan}

\author{T.~Tohyama} 
\affiliation{Yukawa Institute or Theoretical Physics, Kyoto University, Kyoto 606-8502, Japan}

\author{S.~Shin} 
\affiliation{ISSP, University of Tokyo, Kashiwa, Chiba 277-8581, Japan}

\date{\today}

\maketitle

{\bf In contrast to a complex feature of antinodal state, suffering from competing order(s),
the ``pure" pairing gap of cuprates is detected  in the nodal region, which therefore holds the key to the superconducting mechanism. 
The pairing gap has been viewed to be rather conventional, 
closing at the superconducting transition temperature ($T_{\rm c}$). 
However,  the density of states contributed  from the nodal region was claimed to have a gap-like structure even above $T_c$. 
Here we present a missing evidence for a single-particle gap near the node signifying the realization of a phase incoherent $d$-wave superconductivity  above $T_c$ in the optimally doped Bi$_2$Sr$_2$CaCu$_2$O$_{8+\delta}$.
We find that the pair formation is formulated by momentum-independent temperature evolutions of three parameters: 
 a BCS-type energy gap ($\Delta$), a single particle scattering rate (${\Gamma _{{\rm{single}}}}$) and a pair breaking rate (${\Gamma _{{\rm{pair}}}}$). The superconductivity occurs when the ${\Gamma _{{\rm{pair}}}}$ value is suppressed smaller than ${\Gamma _{{\rm{single}}}}$ on cooling, and  the  magnitude of $T_{\rm c}$ in cuprates is governed by a condition of ${\Gamma _{{\rm{single}}}}(T_c)$= ${\Gamma _{{\rm{pair}}}}(T_c)$. 
}

In cuprates, the energy gap (pseudogap) starts opening at a temperature much higher than $T_c$, in some cases above room temperature. 
Many experimental evidences  
\cite{Kondo_competition,Khasanov,Chang_NP,Hashimoto,Eric,Shen_PNAS,Yazdani_PG,He_science,Kondo_MDC} 
point to a competing-order origin, rather than the preformed pair, for the pseudogap observed around the antinode with the maximum energy gap. 
On the other hand, 
the energy gap near the node is expected to open due to the electron pairing as it is free from a contamination by the competing order establishing around the antinode \cite{Kondo_MDC,Shen_BCS,Yazdani}.  
Unveiling the nature of the spectral gap near the node is therefore crucial to elucidate the superconducting mechanism in cuprates.
A difficulty however is the small magnitude of the gap, which has been challenging the experimentalists to investigate.

It has been proposed that the pairing-gap evolution with temperature simply follows the conventional BCS function \cite{Shen_BCS}, and   
 Fermi arcs (disconnected segments of gapless Fermi surface) \cite{Norman_arc} emerge at $T_c$ \cite{Shen_BCS,Yazdani,Kanigel_NP,Kanigel_PRL,Nakayama,Shen_PNAS}, marking momentum borders between the superconducting and the competing pseudogap regions \cite{Shen_BCS,Shen_PNAS}. 
However, it seems contradicting the observations of the Nernst and diamagnetic effects above $T_c$ \cite{Nernst}, which are viewed as signatures of a phase incoherent superconductivity. Notably, the transport properties are sensitive to the low energy excitations near the node,  thus a further study is required.

Recently, a contrasting view was proposed \cite{Dessau_NP,Dessau_PRB,Kondo_MDC}:
its underlying idea is that one should discard the notion of electron quasiparticles, instead pay attention to the density of states, which 
is an effective way of judging the existence of energy gap. 
Accordingly a momentum integration of angle-resolved photoemission spectroscopy (ARPES)
spectra has been performed over a selected part of the momentum space.
This quantity contributed from the nodal region was found to have a gap-like structure even above $T_c$ \cite{Dessau_NP,Dessau_PRB}. 
The result seems to be in direct opposition to the above well-accepted view.
Nevertheless, a direct observation of the single-particle gap reaching this conclusion is still missing, and  it is 
 strongly desired in order to facilitate a better understanding of the nodal state above $T_c$. 

Here we examine the momentum-resolved spectra obtained 
 by a laser ARPES \cite{Kiss}. 
 The ultra-high energy resolution and bulk-sensitivity achieved by utilizing a low-energy laser source ($h\nu$=7eV) enabled us to obtain high quality spectra with an extremely sharp line shape.  
We demonstrates, within the quasiparticle picture, an isotropic temperature evolution of a point nodal pair formation 
persistent way above $T_c$ in cuprates (see Fig.4). 
We find that 
the magnitude of $T_c$ is determined by a mutual relation between  
 the single particle scattering rate (${\Gamma _{{\rm{single}}}}$) and  pair breaking rate (${\Gamma _{{\rm{pair}}}}$), which are both required for reproducing the ARPES spectra, and it is described as  ${\Gamma _{{\rm{single}}}}(T_c)$= ${\Gamma _{{\rm{pair}}}}(T_c)$. Importantly the previously introduced measure (momentum-integrated spectrum) is not capable of separating these 
 two quantities,  and thus the examination of one-particle spectra addressed in this study would be essential to formulate the pairing mechanism of cuperates. 
 
In Fig.1, we show typical data obtained inside the nodal region where the Fermi arc (bold orange curve in the inset of Fig.1d) was previously claimed to appear at $T_c$ \cite{Yazdani,Shen_BCS,Kanigel_NP,Kanigel_PRL,Shen_PNAS}. 
The ARPES intensity map divided by the Fermi function  (see Fig.1a) shows an energy gap and  
an upper branch of the Bogoliubov dispersion at low temperatures, as an indication of the pairing state.
We extract the spectra
at the Fermi momentum ($k_{\rm F}$) over a wide range of temperature in Fig.1b, and plot the peak energies ($\varepsilon _{\rm peak}$s) in Fig.1e. 
  In the same panel, we also plot $\varepsilon _{\rm peak}$s of EDCs symmetrized about the Fermi energy ($E_{\rm F}$) to remove the effect of Fermi cut-off \cite{Norman_arc}, and confirm a consistency between the two different methods. 
The obtained ${\varepsilon _{\rm peak}}(T)$ clearly disagrees with the BCS gap evolution (blue solid curve in Fig.1e) since the gap is open even at $T_c$ (=92K) (see green spectra in Fig.1b and 1c). We also find another abnormal feature:  two peaks in the spectra abruptly merge to one peak  with increasing temperature above $T_c$.
 Even if assuming a phase fluctuation slightly above $T_c$, still such a BCS-type curve (blue dashed curve) does not  fit to the data.
   
To pin down the cause of this anomaly, we examine the momentum variation of ${\varepsilon _{\rm peak}}(T)$ in Fig.2a. 
Surprisingly,  the gap does not close at $T_c$ regardless of $k_{\rm F}$ points. 
The  symmetrized EDCs for various $k_{\rm F}$s   (Figs. 2e and 2f)  clearly demonstrate that the $d$-wave gap with a point node persists at  $T_c$ (Fig.2h),
providing an evidence for absence of Fermi arc at the superconducting phase transition. 
It is further confirmed in Fermi-function divided band dispersions (Fig.2d) measured at $T_c$ along several momentum cuts (color lines in Fig.2g).   
The loss of spectral weight at $E_{\rm F}$ due to the gap opening is seen for all the maps except for at the node (see supplemental Fig.S5 for more details). 
Our high resolution data also show other inconsistencies with the previous expectations \cite{Kanigel_NP,Nakayama,Shen_BCS,Shen_PNAS}. 
First, the length of arc with a single spectral-peak  (${\varepsilon _{\rm peak}}=0$) is off the line crossing the origin against the nodal liquid behavior (Fig.2c). 
Secondly, the temperature evolution of such an arc is gradual up to way above $T_c$ with no 
 indication of borders separating two distinct states.

For a further investigation, we normalize each curve of ${\varepsilon _{\rm peak}}(T)$ to the maximum value at the lowest temperature in Fig.2b.
One can confirm that the data is mismatched with the conventional BCS curve (a red dashed curve) even in the close vicinity of the node.
More importantly, the decreasing behavior of ${\varepsilon _{\rm peak}}(T)$ down to zero  becomes more gradual 
with getting away from the node, and  it eventually follows a BCS-type gap function 
with an onset at 135K (a green curve). 
Here we point out that the peak energy of a spectrum  underestimates the ``real" energy gap ($\varepsilon _{\rm peak}<\Delta$), when the  peak width becomes larger than $\Delta$ as discussed elsewhere \cite{Varma,Chubukov,Kondo_MDC} and simulated in supplemental Fig.S6.
This situation, in fact, occurs at high temperatures, and it gets more serious toward the node with smaller $\Delta$. 
The characteristic  momentum variation of ${\varepsilon _{\rm peak}}(T)$ in Figs. 2a and 2b, therefore, could be a natural consequence of $\Delta (T)$ having the same onset temperature for all  directions ($\phi$s). 
 We find below that a model spectral function, $\pi A({k_F},\omega ) = \Sigma ''/[{(\omega  - \Sigma ')^2} + {{\Sigma ''}^2}]$, with such a BCS-type $\Delta(T,\phi)$ indeed reproduces the ARPES spectra, whereas the traditionally used assumption of $\Delta (T,\phi) \equiv {\varepsilon _{\rm{peak}}}(T,\phi)$ is invalid.

The self-energy ($\Sigma  = \Sigma ' + i\Sigma ''$) we use has a minimal representation with 
 two different scattering rates: single  particle scattering rate ($\Gamma_{\rm single}$) and pair breaking rate ($\Gamma_{\rm pair}$) \cite{Norman_PRB,Chubukov},
\begin{equation} \label{self}
\Sigma ({k_F},\omega ) =  - i{\Gamma _{{\rm{single}}}} + {\Delta ^2}/[\omega  + i{\Gamma _{{\rm{pair}}}}].
\end{equation}
$\Gamma_{\rm single}$ causes the broadening of a peak width, and $\Gamma_{\rm pair}$ fills the spectral weight around $E_{\rm F}$.
We emphasize that the intensity at $E_{\rm F}$ in a gapped spectrum becomes non-zero only when ${\Gamma _{{\rm{pair}}}} \ne 0$ as simulated in supplemental Fig.S6a.
Our spectra measured at the low temperatures ($T \ll {T_c}$) have a negligible intensity at $E_{\rm F}$, which 
 ensures that our data are almost free from  impurity-causing pair breaking effect.
At elevated temperatures, we observe a remarkable gap filling (see Fig.1d and supplemental Fig.S3). 
Significantly, it actually begins from deep below $T_c$, which is not expected in a conventional BCS superconductor. 
Since the data were measured at the extremely high energy resolution ($\Delta\varepsilon$ = 1.4 meV), we can rule out the possibility, assumed before with setting ${\Gamma _{{\rm{pair}}}} \equiv 0$ \cite{Shen_BCS,Shen_PNAS},
that the filling is caused by a spectral broadening due to the experimental energy resolution.
The intensity at $E_{\rm F}$ should instead be a signature of intrinsic pair breaking, hence it must be taken into account for the gap estimation. 
In passing, we note that the $\Gamma_{\rm single}$ and $\Gamma_{\rm pair}$ both  equally increases the intensity around 
$E_{\rm F}$ of the momentum-integrated spectrum previously studied \cite{Dessau_NP,Dessau_PRB, Kondo_MDC} (see supplemental Fig.S7), which is therefore incapable of disentangling these two different scattering rates. 

Following this consideration, we set  ${\Gamma _{{\rm{single}}}}$ and   ${\Gamma _{{\rm{pair}}}}$  to be independent free parameters in Eq.(1). 
First,  we performed a spectral fitting to our ARPES data, assuming $\Delta (T) \equiv {\varepsilon _{{\text{peak}}}}(T)$, which is the traditional way of gap estimation (see the results in supplemental Fig.S8a). 
The obtained parameter of ${\Gamma _{{\rm{single}}}}(T)$ (middle panel of Fig.S8a) is strongly deviated from a monotonic decrease on cooling, having an unrealistic upturn around the temperature at which  the ${\varepsilon _{\rm{pair}}}$ becomes zero. It is inconsistent with a rather isotropic scattering mechanism observed around the node (see supplemental Fig.S9), indicating that 
 the spectrum with a single peak (${\varepsilon _{\rm{pair}}}=0$) can have an energy gap ($\Delta \ne 0$), and  such a spectral width  overestimates the scattering rate. 
 
 We found that this circumstance is corrected by applying 
a BCS-type gap function with an onset at 135K (a green curve in Fig.2b and Fig.3b) for all $\phi$s.
In Fig.3c, we fit  Eq.(1) with such a gap function $\Delta$ 
to our ARPES data near the node measured over a wide temperature range.
The fitting curves (red  curves) almost perfectly reproduce the data (black  curves) for all the $k_{\rm F}$ points and temperatures. 
The obtained ${\Gamma _{{\rm{single}}}}(T)$  (Fig.3b) in the gapped region agrees with the reliable values for the node, which can be determined simply from the spectral width.
The obtained ${\Gamma _{{\rm{pair}}}}(T)$ curves (Fig.3b)  are also almost identical  for all the $\phi$s.
The consistency in our results
pointing to the isotropic scattering mechanism
validates our model spectral function characterized by the BCS-type $\Delta(T,\phi)$. 
The applied onset temperature, 135K,  is almost the same as 
that of Nernst and diamagnetic effects ($\sim 125$K) \cite{Nernst}, which are viewed as signatures of phase-incoherent superconductivity. A comparable value of the pairing temperature ($T_{\rm pair}$) is 
also estimated from other spectroscopic techniques \cite{Kondo_pair,Yazdani_pair}.
Therefore, we assign 135K to be the $T_{\rm pair}$  of our samples. 
This is further supported by the  signature of pairing  seen in 
the behavior of ${\Gamma _{{\rm{single}}}}(T)$ (Fig.3b): 
the decrease of its value upon cooling is accelerated across $T_{\rm pair}$, showing a deviation from the linear behavior \cite{barivsic2013universal}. 

The relation between ${\Gamma _{{\rm{pair}}}}$ and ${\Gamma _{{\rm{single}}}}$  is expected to provide rich information.
Intriguingly, the superconductivity occurs when the magnitude of ${\Gamma _{{\rm {pair}}}}$ is reduced smaller than that of  ${\Gamma _{{\rm {single}}}}$; 
the $T_c$  is coincident with the temperature at which ${\Gamma _{{\rm{single}}}}(T)$ and  ${\Gamma _{{\rm{pair}}}}(T)$ crosses (a magenta circle in Fig.3b).     
The magnitude of $T_{\rm pair}$  is reported to be comparable among different cuprate families (120K$\sim$150K) \cite{Kondo_pair} with significantly different $T_c$s.
Furthermore, the ${\Gamma _{{\rm{single}}}}(T)$ also seems to be less sensitive to the different compounds \cite{kondo_kink,Zhou}. Therefore, 
the pair breaking effect, which controls the  fulfillment of ${\Gamma _{{\rm {pair}}}} < {\Gamma _{{\rm {single}}}}$, 
 is predicted to be a critical factor determining the $T_c$ value of cuprates.
 Notably, a remarkable difference in filling behaviors of the spectral gap is indeed observed between  Bi2212 and Bi$_2$Sr$_2$CuO$_{6+\delta}$ with about three times different $T_c$s \cite{Kondo_pair,Kondo_MDC}.

We summarize our conclusion in Fig.4 by drawing a schematic temperature evolution of the pairing gap. 
A  phase-incoherent, point nodal $d$-wave superconductivity occurs below $T_{\rm pair} \sim 1.5T_c$ (Figs. 4b and 4c).
The anomalous momentum variation of ${\varepsilon _{\rm peak}}(T)$ (Fig .2a) is attributed to  
the  pair breaking effect on the point nodal state, which is facilitated at elevated temperatures.
To fully understand the present results, insight of the spacially inhomogeneous state \cite{Pan}
might be essential. However, only that cannot explain our data, since 
the local density of states itself seems to have the behavior of gap filling at $E_{\rm F}$ with temperature \cite{Yazdani_gamma}. 
The  competing nature of pseudogap state evolving around the antinode \cite{Kondo_competition,Chang_NP} is a plausible cause for 
the unique scattering mechanism causing the significant suppression of $T_c$ from $T_{\rm pair}$. 
To evaluate this speculation, however, the more detailed theoretical inputs are required. 

\textbf{Methods} 

Optimally doped Bi$_2$Sr$_2$CaCu$_2$O$_{8+\delta}$ (Bi2212) single crystals with $T_{\rm c}$=92K 
were grown by the conventional floating-zone (FZ) technique. 
A sharp superconducting transition with a width less than 1 K was confirmed (see supplemental Fig.S1).  
ARPES data was accumulated using a laboratory-based system consisting of a Scienta R4000 electron analyzer and  a 6.994 eV laser. The overall energy resolution in the ARPES experiment was set to 1.4 meV for all the measurements.   

In order to accomplish the temperature scan of spectra at a high precision, 
we applied a technique of the local sample heating, 
which thermally isolates the sample holder with a heat switch from the lest of the system at elevated temperatures. 
It minimizes the degassing, allowing us to keep the chamber pressure better than $2 \times {10^{ - 11}}$ torr
during the entire temperature sweeping; no sample aging was confirmed (supplemental Fig.S4). 
This method also prevents the thermal expansion of sample manipulator, and it enables us to
take data in fine temperature steps with automated measurement of temperature scan from precisely the same spot on the crystal surface, which were essential to achieve the aim of the present study.
 
 {\bf  Acknowledgements}\\
  We thank M. Imada, S. Sakai and T. Misawa for useful discussions. This work is supported by JSPS (KAKENHI Grants No. 24740218 and FIRST Program).


\clearpage
\begin{figure*}
\includegraphics[width=6.5in]{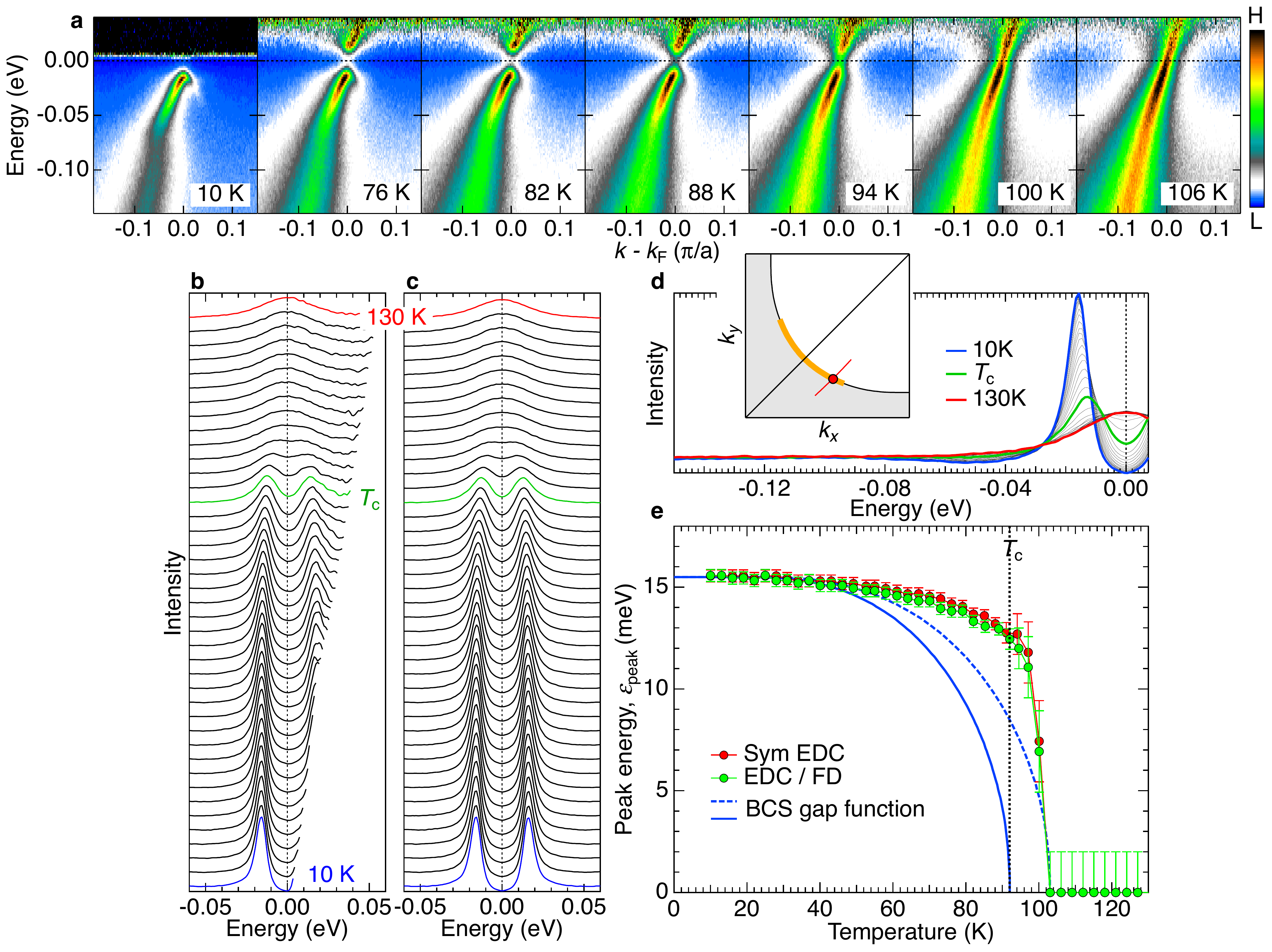}
\caption{ {\bf Temperature evolution of ARPES spectra in the nodal region.}
 {\bf a}, Dispersion maps at several temperatures measured along a momentum cut (a red line in the inset of {\bf d}). 
Each map is divided by the Fermi function at the measured temperature. 
{\bf b}, Temperate evolution of EDCs at $k_{\rm F}$ (a circle in the inset of {\bf d}) from deep below  (10K) to much higher than $T_c$ (130K). 
Each spectrum is divided by the Fermi function at the measured temperature. 
{\bf c}, The same data as in {\bf b}, but symmetrized about $E_{\rm F}$. 
{\bf d}, The same data as in {\bf c}, but plotted without an offset. The inset represents the Fermi surface.
The bold orange line indicates the momentum region where the Fermi arc was previously claimed  to emerge at $T_c$.  
{\bf e}, Peak energies of spectra in {\bf b} and {\bf c} plotted as a function of temperature, ${\varepsilon _{\rm peak}}(T)$. 
The  solid and dashed blue curves show the BCS gap function with an onset at $T_c$ (92K) and slightly above $T_c$, respectively. } 
\label{fig1}
\end{figure*}

\clearpage
\begin{figure*}
\includegraphics[width=6.5in]{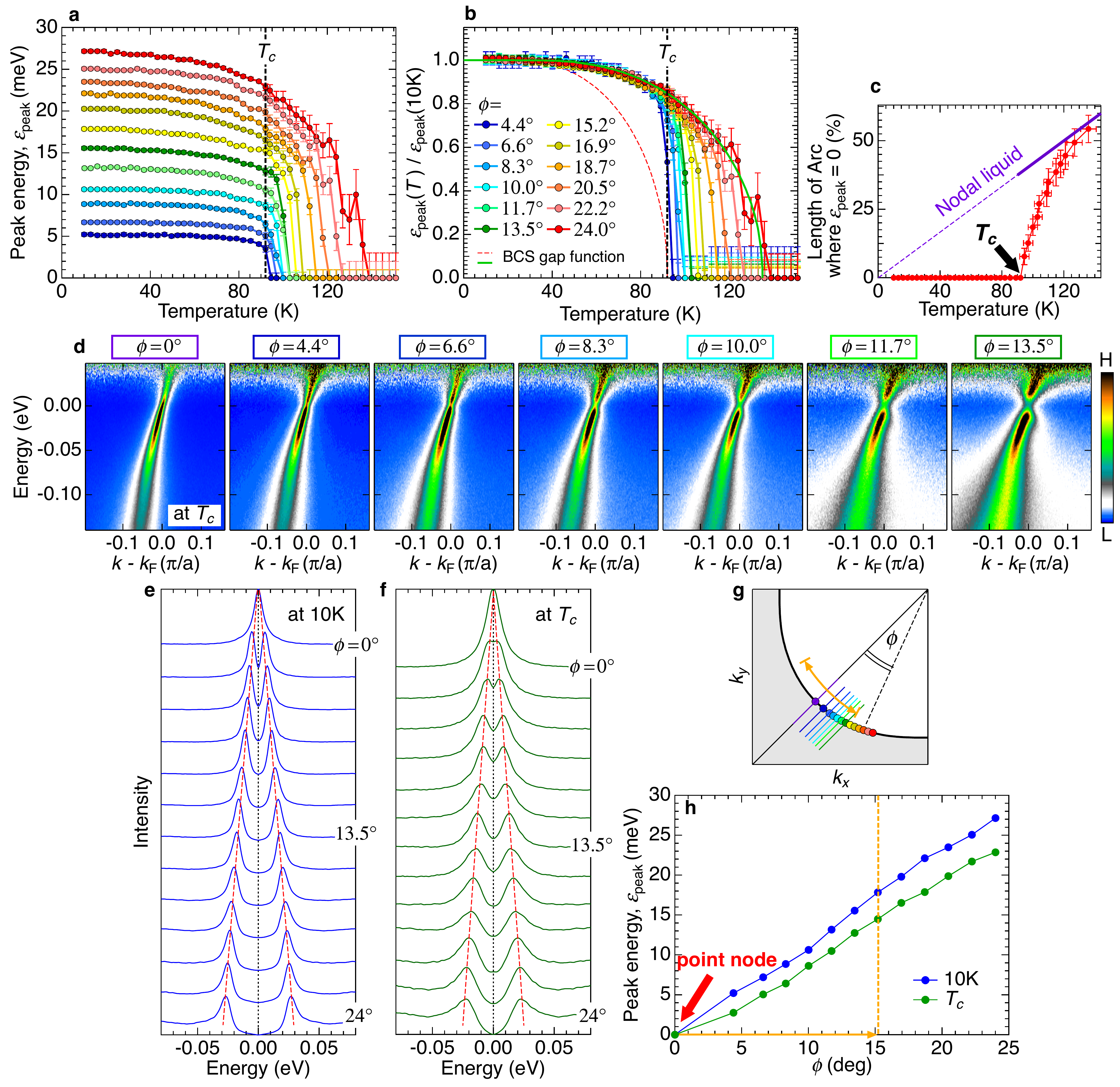}
\caption{}
\label{fig1}
\end{figure*}

\clearpage
 FIG. 2:   {\bf Momentum variation of spectra and the absence of  Fermi arc at the superconducting transition.  }
{\bf a}, Temperature dependence of spectral peak energy, ${\varepsilon _{\rm peak}}(T)$, at various $k_{\rm F}$ points (circles in {\bf g}). 
{\bf b}, The same data as in {\bf a}, but normalized to the maximum value at the lowest temperature. 
A red dashed curve  and a green solid curve are the BCS gap function with an onset at $T_c$ (92K) and $T_{\rm pair}$ (135K), respectively.
{\bf c},  Length of arc, at which the single-particle spectra at $k_{\rm F}$ points have single peaks  (${\varepsilon _{\rm peak}}=0$).
The proposed behavior of nodal liquid \cite{Kanigel_NP,Nakayama} is also plotted: the arc length above $T_c$ is linear in $T$, and it vanishes if extrapolated to $T=0$.
{\bf d}, Dispersion maps at $T_c$  along several momentum cuts  (color lines in {\bf g}). 
Each map is divided by the Fermi function at the measured temperature.  
The described $\phi$ is the direction of $k_{\rm F}$ point (defined in {\bf g}). 
{\bf e}, {\bf f},  Symmetrized EDCs at $k_{\rm F}$ 
over a wide range of angle $\phi$ (circles in {\bf g}) at 10K and $T_c$ (=92K), respectively.
The red dotted lines are guide to the eyes for the gap evolution.
 {\bf g},  Fermi surface.
{\bf h}, Fermi angle $\phi$ dependence of ${\varepsilon _{\rm peak}}$(10K) and ${\varepsilon _{\rm peak}}(T_c)$. The orange arrows  in {\bf g} and {\bf h} indicate 
the momentum region where the Fermi arc was previously claimed  to emerge at $T_c$.

\clearpage

\begin{figure*}
\includegraphics[width=6in]{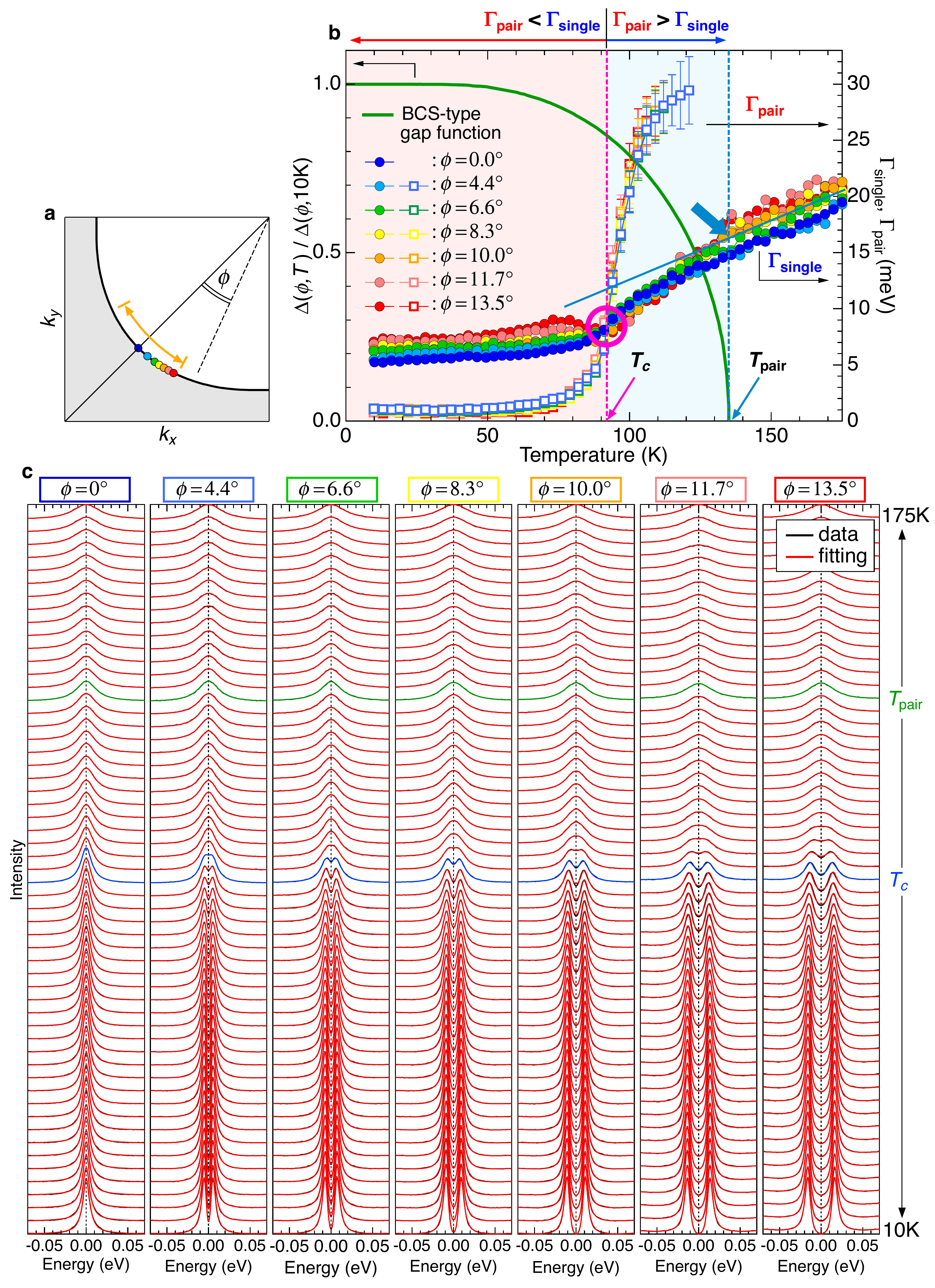}
\caption{} 
\label{fig1}
\end{figure*}

\clearpage
 FIG. 3:  {\bf ARPES spectra reproduced by a minimal model spectral function.}
{\bf a},  Fermi surface. 
The orange arrow indicates the momentum region where the Fermi arc was previously claimed to emerge at $T_c$.
{\bf b}, The BCS-type gap function  used for the fitting (a green curve), and the obtained 
single particle scattering rate ($\Gamma_{\rm single}$) and the pair breaking rate ($\Gamma_{\rm pair}$) in Eq.(1). 
The values of $\Gamma_{\rm pair}$ at high temperatures are not plotted, since the spectral shape is insensitive to the $\Gamma_{\rm pair}$  when 
$\Delta$
is small or zero, and thus it is impossible to  determine the value.
A small hump seen in the $\Gamma_{\rm pair}$ around 75K for $\phi  = 11.7^\circ$ and $13.5^\circ$ comes from a slight difficulty of fitting to the spectra with a peak-dip-hump shape due to the mode coupling, which appears below $T_c$ and gets pronounced with approaching the antinode.  Magenta circle marks the crossing point of $\Gamma_{\rm single}(T)$ and $\Gamma_{\rm pair}(T)$. Large arrow indicates the temperature, at which the  $\Gamma_{\rm single}(T)$ deviates from a $T$-linear behavior upon cooling.
{\bf c}, ARPES spectra (black curves) and fitting results (red curves) providing the parameters in {\bf b}. 
We added a small background linear in energy ($\propto \left| \omega  \right|$) to the fitting function $A({k_F},\omega )$ for $\phi=11.7^{\circ}$ and $13.5^{\circ}$, 
in order to properly extract the scattering rates.
All the fitting curves are convoluted with a gaussian that has the width of  the experimental energy resolution ($\Delta \varepsilon  = 1.4$meV). Nevertheless, note that the $\Delta \varepsilon$ value  is so small  that the difference in shape between the two with and without the convolution is negligible.

\clearpage
\begin{figure*}
\includegraphics[width=6.5in]{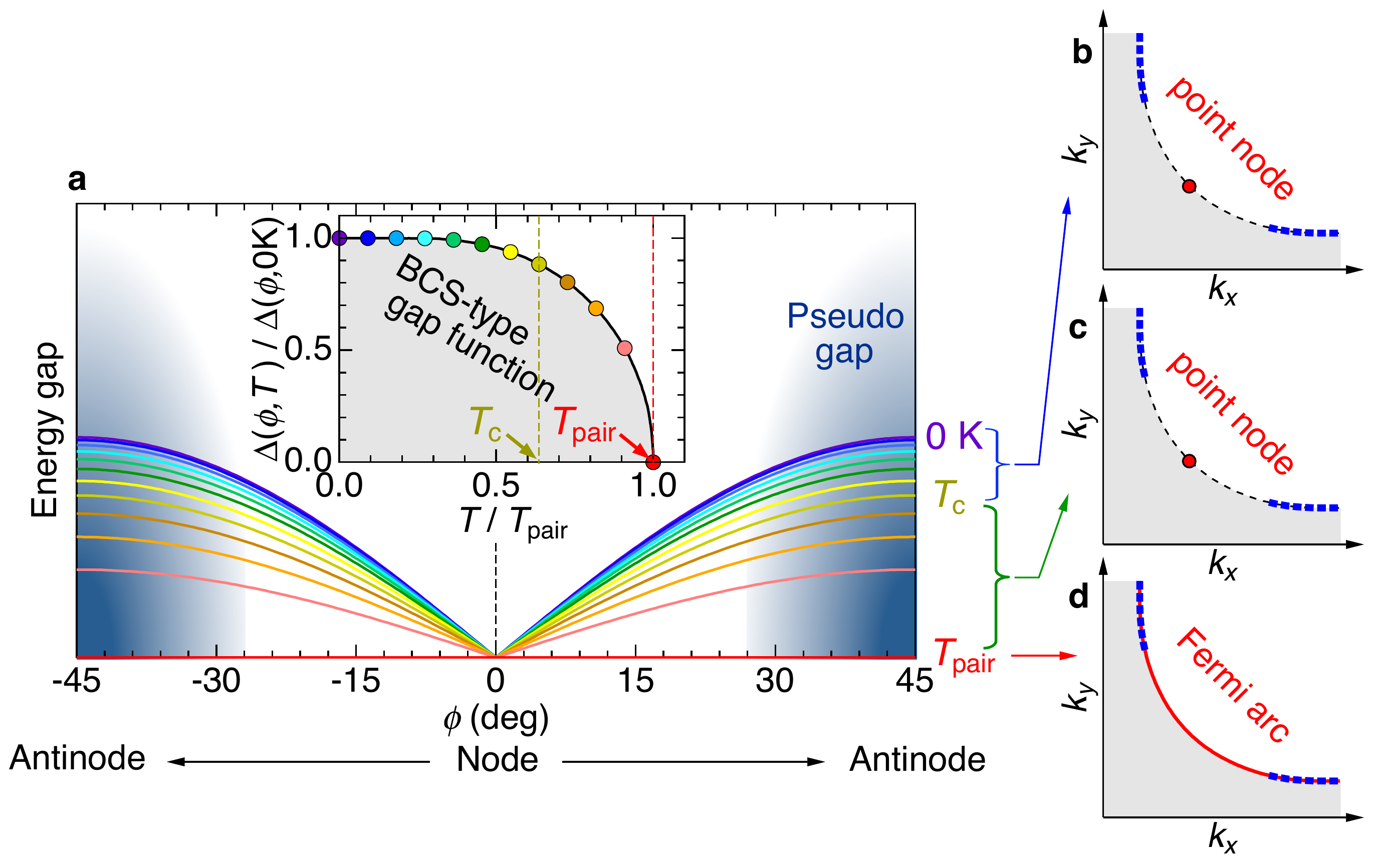}
\caption{ {\bf  Schematic pairing-gap evolution based on our ARPES results. }
{\bf a}, Temperature dependence of $d$-wave pairing gap with BCS-type temperature dependence (inset curve) regardless of directions ($\phi$s) along the Fermi surface. Temperatures for each curve are indicated in the inset with colored circles. The gapped Fermi surface with a point node  ({\bf b}) below $T_c$ 
persists ({\bf c})  beyond $T_c$  up to the temperature of pair formation ($T_{\rm pair}$). 
{\bf d},  Emergence of the gapless Fermi arc centered at the node due to the pseudogap  evolution around the antinode \cite{Kondo_MDC,Kohsaka}; 
while the antinodal region is not observable at the low photon energies as 7eV, 
the studies with higher energy photons  demonstrate that the competing pseudogap state emerges  at
$\left| \phi  \right| > 25^\circ$ \cite{Kondo_MDC}.
 } 
\label{fig1}
\end{figure*}

\end{document}


\title{Supplementary information for\\
Observation of quasiparticles for phase incoherent $d$-wave pairing in Bi2212}

\author{Takeshi Kondo}
\affiliation{ISSP, University of Tokyo, Kashiwa, Chiba 277-8581, Japan}

\author{W.~Malaeb} 
\affiliation{ISSP, University of Tokyo, Kashiwa, Chiba 277-8581, Japan}

\author{Y.~Ishida} 
\affiliation{ISSP, University of Tokyo, Kashiwa, Chiba 277-8581, Japan}

\author{T.~Sasagawa} 
\affiliation{Materials and Structures Laboratory, Tokyo Institute of Technology, Yokohama, Kanagawa 226-8503, Japan}

\author{T.~Tohyama} 
\affiliation{Yukawa Institute for Theoretical Physics, Kyoto University, Kyoto 606-8502, Japan}

\author{S.~Shin} 
\affiliation{ISSP, University of Tokyo, Kashiwa, Chiba 277-8581, Japan}

\date{\today}

\maketitle

\section {Samples and Experimental method}
Optimally doped Bi$_2$Sr$_2$CaCu$_2$O$_{8+\delta}$ (Bi2212) single crystals with $T_{\rm c}$=92K 
were grown by the conventional floating-zone (FZ) technique. 
Magnetic susceptibility for the single crystal is shown in Fig.\ref{SQUID}. 
A sharp superconducting transition with a width less than 1 K, indicative of a high quality, is confirmed.  
ARPES data was accumulated using a laboratory-based system consisting of a Scienta R4000 electron analyzer and  a 6.994 eV laser (the 6th harmonic of Nd:YVO4 quasi-continuous wave with a repetition rate of 240 MHz). The overall energy resolution in the ARPES experiment was set to 1.4 meV for all the measurements.
   
In order to accomplish the temperature scan of spectra at a high precision, 
we made a special sample stage;
we use a heat switch to thermally isolate the sample holder with a heater (totally $20 \times 30 \times 10$ mm size)  
from the lest of the system at elevated temperatures. 
With this technique, the samples are efficiently heated up  from 10K up to 300K 
while keeping the temperature of manipulator rod other than the sample stage lower than 35K. 
It minimizes the degassing  
and keeps the  pressure in the measurement chamber better than $2 \times {10^{ - 11}}$ torr
during the entire temperature sweeping. 
The method of local heating also prevents the thermal expansion of the $long$ manipulator rod, and thus the sample position becomes unchanged with temperature. 
It enabled us to
take data in fine temperature steps with the automated measurement of temperature scan from precisely the same spot on the crystal surface, which were essential to achieve the aim of this study.

In Fig.\ref{WithOffset} and \ref{WithoutOffset}, we show the ARPES spectra obtained by the temperature sweeping for  
various $k_{\rm F}$ points with and without an offset, respectively. These are the original data 
used for the analysis in the main paper. 
The ARPES technique is sensitive to the condition of sample surface, and the biggest challenge in it is to 
prevent the aging of sample surface with time. 
In Fig.\ref{AgingCheck}, we carefully check the stability of sample surface in our experiment 
by comparing two sets of spectra measured before and after the temperature scan for all the momentum cuts we measured. 
While it took about 12 hours to complete each temperature sweeping, 
 almost perfect reproducibility of spectral line shapes is confirmed for all the data.
This ensures that our data measured with the new heating technique 
 are reliable and suitable for conducing the detailed analysis presented in the paper. 
 
 \section {Further Evidence for the absence of the Fermi arc at $T_c$} 
 In the main paper, we demonstrated that the $d$-wave gap with a point node persists at $T_c$ by plotting the symmetrized EDCs at $k_F$ points (see Fig.2f).  Here we demonstrate that the conclusion is robust and it is not sensitive to the selection of $k_F$ value. 
 Figure \ref{Pointnode}a shows  the Fermi-function divided band dispersions measured at $T_c$ along several momentum cuts (color lines in Fig.\ref{Pointnode}c), which is same as Fig.2d in the main paper. The EDCs corresponding to each panel in Fig.\ref{Pointnode}a are plotted in Fig.\ref{Pointnode}b. 
Along the node (left panel), the spectrum with one peak crosses $E_F$ due to the gap node. 
In contrast, two peak structure (marked with arrows), signifying a gap existence, is observed at $k_F$ (painted with red) for the other directions. This is a direct evidence that the $d$-wave gap with a point node persists, thus the Fermi arc is absent at $T_c$.

\section {Effect of scattering rates, $\Gamma_{\rm single}$ and $\Gamma_{\rm pair}$, on the spectral line shape} 

In the analysis, we used the spectral function with a phenomenological self-energy (ref. \cite{Norman_PRB}) given by  
\renewcommand {\theequation}{S\arabic{equation}}
\begin{eqnarray} \label{self_Norman}
A(k,\omega ) = \frac{1}{\pi } \cdot \frac{{\Sigma ''(k,\omega )}}{{{{\left[ {\omega  - \varepsilon (k) - \Sigma '(k,\omega )} \right]}^2} + \Sigma ''{{(k,\omega )}^2}}},\nonumber \\ \nonumber \\
\Sigma (k,\omega ) =  - i{\Gamma _{{\rm{single}}}} + {\Delta ^2}/[\omega  + \varepsilon (k) + i{\Gamma _{{\rm{pair}}}}].
\end{eqnarray}  
Here $\Delta$ is the energy gap, and  $\varepsilon (k)$ the energy dispersion with $\varepsilon (k_F)=0$. $\Gamma_{\rm single}$ and $\Gamma_{\rm pair}$ are the single particle scattering rate and the pair breaking rate, respectively.
In Fig.\ref{Simulation}, we examine the effect of these scattering rates 
on the spectral line shape at $k_F$ with fixing $\Delta$  to 10 meV.
To make it realistic, 
all the curves plotted are convoluted with a gaussian that has the width of  the experimental energy resolution ($\Delta \varepsilon  = 1.4$meV). Nevertheless, note that the $\Delta \varepsilon$ value  is so small  that the difference in  shape from original curves is negligible.  
Figure \ref{Simulation}a shows the $\Gamma_{\rm single}$ dependence of the spectral function when  the $\Gamma_{\rm pair}$ is fixed to zero.
One finds that the intensity at $E_{\rm F}$ is always zero regardless of the $\Gamma_{\rm single}$ value, whereas the spectral width gets broadened with increasing $\Gamma_{\rm single}$. 
This situation strongly contrasts to that in our ARPES data (Fig.\ref{WithoutOffset}), which clearly show the filling of spectral weight at $E_{\rm F}$ with increasing temperature. The behavior of gap filling can be produced by inputting non-zero $\Gamma_{\rm pair}$ values in Eq.(\ref{self_Norman}). It is demonstrated in 
Fig.\ref{Simulation}b, where the spectra with several $\Gamma_{\rm pair}$ values are calculated with fixing $\Gamma_{\rm single}$ to zero.
 Importantly, the energy position of spectral peak (${\varepsilon _{{\text{peak}}}}$) dramatically decreases with an increase of the $\Gamma_{\rm pair}$ value as a consequence of the spectral filling around $E_{\rm F}$, and eventually it becomes zero. This simulation indicates that the peak energy underestimates the ``real" energy gap (${\varepsilon _{{\text{peak}}}} < \Delta $) when the magnitude of $\Gamma_{\rm pair}$ becomes large. 
 
Here we should point out that the following self-energy has been used to estimate the magnitude of $\Delta$ by a fitting to the ARPRS spectra in the many previous reports,
\renewcommand {\theequation}{S\arabic{equation}}
\begin{eqnarray} \label{self_simple}
\Sigma ({k_F},\omega ) \equiv  - i{\Gamma _{{\text{single}}}} + {\Delta ^2}/\omega.
\end{eqnarray}  
This is the same formula as Eq.(\ref{self_Norman}) setting ${\Gamma _{{\text{pair}}}} = 0$, and thus
it always provides one the spectral intensity of zero at $E_{\rm F}$ as discussed above (see Fig.\ref{Simulation}a). 
We emphasize that the spectral weight around  $E_{\rm F}$ observed in the ARPES data was previously attributed to the spectral broadening coming from the finite energy resolution in ARPES (for example, see the supplementary information of ref.\cite{Shen_BCS}). 
The extremely high energy resolution in our equipment enabled us to reveal 
that the spectral weight at $E_{\rm F}$ is not an artifact 
due to the experimental resolution, but an intrinsic signature of the pair breaking effect.
This signifies that the  Eq.(\ref{self_Norman}) is more suitable than Eq.(\ref{self_simple}) as the model spectral function of cuprates,
and that the effect of $\Gamma_{\rm pair}$ must be taken into account to properly estimate the value of $\Delta$ from the spectral shape.  

In Fig.\ref{Simulation}c, we simulate a special case setting 
  $\Gamma_{\rm pair} =\Gamma_{\rm single}$ in Eq.(\ref{self_Norman}).
  The self-energy and the corresponding spectral function at $k_F$ under this condition are derived as follow;
  \renewcommand {\theequation}{S\arabic{equation}}
\begin{eqnarray} \label{BCS_spectrum}
\Sigma ({k_F},\omega ) =  - i{\Gamma} + {\Delta ^2}/[\omega  + i{\Gamma}],
\nonumber \\ \nonumber \\
A({k_F},\omega ) = {1 \over 2}\left[ {{\Gamma  \over {{{\left( {\omega  - \Delta } \right)}^2} + {\Gamma ^2}}} + {\Gamma  \over {{{\left( {\omega  + \Delta } \right)}^2} + {\Gamma ^2}}}} \right],
\end{eqnarray}  
where $\Gamma  \equiv {\Gamma _{{\text{single}}}} = {\Gamma _{{\text{pair}}}}$.
This formula, consisting of double Lorentzian functions, is  familiar as the phenomenological BCS spectral function \cite{Schrieffer,JC,Matsui}.
The calculated spectra for various $\Gamma$ values are plotted in Fig.\ref{Simulation}. One can confirm a couple of discrepancies in these from our data.  
First, the spectral peak in the ARPRS data has an asymmetric shape about the peak energy, differently from the Lorentzian shape.
Secondly, 
the effect of gap filling in this simulation  is quite sensitive to the magnitude of $\Gamma$, and it is far more drastic than that observed in our data at elevated temperatures.
These lead us to conclude that the Eq.(\ref{BCS_spectrum}) is not a proper formula of spectral function for cuprates either; that is, 
the assumption of $\Gamma_{\rm pair} =\Gamma_{\rm single}$ is not suitable.

We also stress that the Dynes function \cite{Dynes} given by,  
\begin{equation} \label{Dynes} 
{I_{{\rm{DOS}}}}(\omega ) = {\mathop{\rm Re}\nolimits} \left[ {{{\omega  - i\Gamma } \over {\sqrt {{{\left( {\omega  - i\Gamma } \right)}^2} - {\Delta ^2}} }}} \right],
\end{equation} 
is the density-of-states version of Eq.(\ref{BCS_spectrum}) with  $\Gamma  \equiv {\Gamma _{{\text{single}}}} = {\Gamma _{{\text{pair}}}}$, hence it is also not adequate for cuprates.  This is understandable because there is no good physical reason that these two different scattering rates should be equal. 
Our results demonstrate that a proper combination of $\Gamma_{\rm single}$ and $\Gamma_{\rm pair}$ is required to reproduce the data.  

In the previous reports \cite{Dessau_NP,Dessau_PRB,Kondo_MDC}, the spectrum integrated along a selected momentum cut
was utilized to study the gap evolution with temperature and its relation with the electron scattering. 
Here we argue the necessity of investigating the one-particle spectrum, rather than its momentum integration, for such a study. 
In Fig.\ref{DOSsimulation}, we demonstrate that a momentum-integrated spectrum, ${I_{{\rm{DOS}}}}(\omega ) = \int {A(k,\omega )} dk$ for $A(k,\omega )$ of Eq.(\ref{self_Norman}),  is equally sensitive to the $\Gamma_{\rm single}$ value (Figs. \ref{DOSsimulation}{\bf a} and \ref{DOSsimulation}{\bf b})
and the $\Gamma_{\rm pair}$ value (Figs. \ref{DOSsimulation}{\bf c} and \ref{DOSsimulation}{\bf d}); these two scattering rates both fill the spectral intensity around $E_F$, while the gap size $\Delta$ is fixed to a constant value (15 meV).
The spectra of ${I_{{\rm{DOS}}}}(\omega )$ calculated under the condition of ${\Gamma _{{\rm{single}}}} = {\Gamma _{{\rm{pair}}}}$  (Figs.   \ref{DOSsimulation}{\bf e} and  \ref{DOSsimulation}{\bf f}) reproduce those in Figs.  \ref{DOSsimulation}{\bf a} and \ref{DOSsimulation}{\bf b}. As naturally expected, only the half values of ${\Gamma _{{\rm{single}}}}$ and  ${\Gamma _{{\rm{pair}}}}$ 
are sufficient to cause the same effect in this case. 
Notably, the obtained curves are  identical to Dynes function in Eq.(\ref{Dynes}) as pointed out above, which is demonstrated for several values of $\Gamma$ in  Fig.\ref{DOSsimulation}{\bf g}.  These simulations providing identical results (Figs.  \ref{DOSsimulation}{\bf a}, \ref{DOSsimulation}{\bf c}, \ref{DOSsimulation}{\bf e}, and \ref{DOSsimulation}{\bf g})  signify that the 
  momentum-integrated spectrum is not capable of extracting the mutual relationship between the two scattering rates (${\Gamma _{{\rm{single}}}}$ and ${\Gamma _{{\rm{pair}}}}$).

In contrast to it,  the effects of  ${\Gamma _{{\rm{single}}}}$ and ${\Gamma _{{\rm{pair}}}}$ on the line shape of the momentum-resolved spectrum $A(k,\omega )$ are independent, 
broadening the width and filling the weight around $E_F$, respectively.
This circumstance is demonstrated in Fig.\ref{Simulation}, and also
 seen in  Fig.\ref{DOSsimulation}{\bf h}, at which the spectra at $k_F$ are extracted from images in Figs. \ref{DOSsimulation}{\bf b}, \ref{DOSsimulation}{\bf d}, and \ref{DOSsimulation}{\bf f}. (Spectral colors of red, blue, and green are corresponding to those of thick dashed lines on  the images.) 
Therefore, the  precise investigation for the line shape of one-particle spectra is required 
  to separate the  three important parameters (${\Gamma _{{\rm{single}}}}$,  ${\Gamma _{{\rm{pair}}}}$, and $\Delta$) in   Eq.(\ref{self_Norman}). The main challenge in this task, which has been troubling the ARPES people,  
  is to accomplish both of a sufficient energy resolution and an ultra-high statistics in the ARPES spectra with no noise at a time, which are usually conflicting with each other. We have overcome this difficulty by using a Laser ARPES with a unique cold finger (mentioned in "Experimental Method"), which was constructed in our lab. 
We emphasize that the Eq.(\ref{self_Norman}) is a minimal model to reproduce the ARPES data, 
 thus the relation among the three physical parameters, revealed in the present work, provides a key ingredient to formulate the mechanism of the high-$T_c$ superconductivity in cuprates.

\section {Spectral fitting under the assumption of $\Delta (T) \equiv {\varepsilon _{{\text{peak}}}}(T)$} 
In the main paper, we claim that the energy of spectral peak underestimates the energy gap (${\varepsilon _{{\text{peak}}}} < \Delta $) close to the node at high temperatures, at which the spectral peak width becomes larger than the magnitude of the energy gap. 
Here we perform a spectral fitting to our ARPES data with Eq.(\ref{self_Norman}), assuming that  
these two values are identical ($\Delta (T) \equiv {\varepsilon _{{\text{peak}}}}(T)$, see upper panel of Fig.\ref{FittingResult}a), 
and demonstrate that  the extracted  parameter [$\Gamma_{\rm single}$ in Eq.(\ref{self_Norman})] has an unrealistic behavior, reflecting the irrelevance of $\Delta (T) \equiv {\varepsilon _{{\text{peak}}}}(T)$. 
The middle and bottom panels of  Fig.\ref{FittingResult}a plot  
the  $\Gamma_{\rm single}(T)$ and $\Gamma_{\rm pair}(T)$, respectively, obtained by the fitting to the spectra in the momentum region, where the Fermi arc was previously claimed to appear at $T_c$ (see an orange arrow in the inset of Fig.\ref{FittingResult}b). 
 For a comparison, we also plot in Fig.\ref{FittingResult}b the corresponding results for  $\Delta (T)$ of the BCS-like gap function, which are 
presented in the main paper (Fig.3c). 
  The difference between the two fitting-results is very clear; 
the former (Fig.\ref{FittingResult}a) shows an abnormal upturn in the  ${\Gamma _{{\text{single}}}}(T)$ curves on cooling in the gapped momentum region, and it contrasts to 
the latter result  (Fig.\ref{FittingResult}b)  showing almost identical curves in ${\Gamma _{{\text{single}}}}(T)$ for all
$\phi$s. 

The  behavior of the abnormal upturn is interpreted as follows. 
 An energy gap is still open above $T_c$ even though the spectrum has a single peak, and thus the  spectral peak width overestimates the scattering rate.
 The similar situation is simulated in Fig.\ref{Simulation}b.  
 Since the degree of the overestimation gets higher with an increase of the energy gap on cooling, the upturn in the ${\Gamma _{{\text{single}}}}(T)$ appears. 
The upturn  seems to be more enhanced with getting away from the node (see middle panel of Fig.\ref{FittingResult}a).
 This is also expected as the overestimation becomes more serious toward the antinode with a larger energy gap. 
This scenario is further validated by the fact that the upturn starts around $T \sim 130$K, which is determined to be 
the onset temperature of  pair formation in the main paper. 

One might think that the cuprates are known to have 
a significantly anisotropic scattering mechanism, 
thus the situation in Fig.\ref{FittingResult}a with the strong momentum variation could be intrinsic. 
To argue against this, here we demonstrate that the  optimally doped Bi2212, in fact, has an isotropic scattering mechanism around the node. 
Figures \ref{IsotropicScattering}a and \ref{IsotropicScattering}b show the spectra at various $k_{\rm F}$ points  over a wide $\phi$ angle  measured at the lowest temperature ($T$=10K) and above the pairing temperature ($T$=150K), respectively. 
At these temperatures, the value of ${\Gamma _{{\text{single}}}}$ can be  estimated simply from the peak width of spectra,
since the magnitude of  ${\Gamma _{{\text{pair}}}}$ is negligible at 10K, and 
the second term in Eq.(\ref{self_Norman}) with ${\Gamma _{{\text{pair}}}}$  is irrelevant
above the pairing temperature.
The values of ${\Gamma _{{\text{single}}}}$  estimated from the peak width of spectra in Fig.\ref{IsotropicScattering}a and \ref{IsotropicScattering}b
are plotted in  Fig.\ref{IsotropicScattering}d with blue and red circles, respectively.
The magnitude is almost constant over a wide $\phi$ centered at the node for both the temperatures, 
which indicates that the scattering mechanism near the node is isotropic. 
This strongly supports  our assertion that the anisotropic behavior of 
${\Gamma _{{\text{single}}}}$ with an upturn on cooling is an artifact. 
Beyond  $\left| \phi  \right| \approx 15^\circ$, the scattering rate abruptly increases. 
This signifies the evolution of pseudogap, which becomes dominant around the antinode \cite{Kondo_competition,Kondo_pair,Lee_Science,Kohsaka}.


\renewcommand{\thefigure}{S\arabic{figure}}

\newpage

\begin{figure*}  
\includegraphics[width=5in]{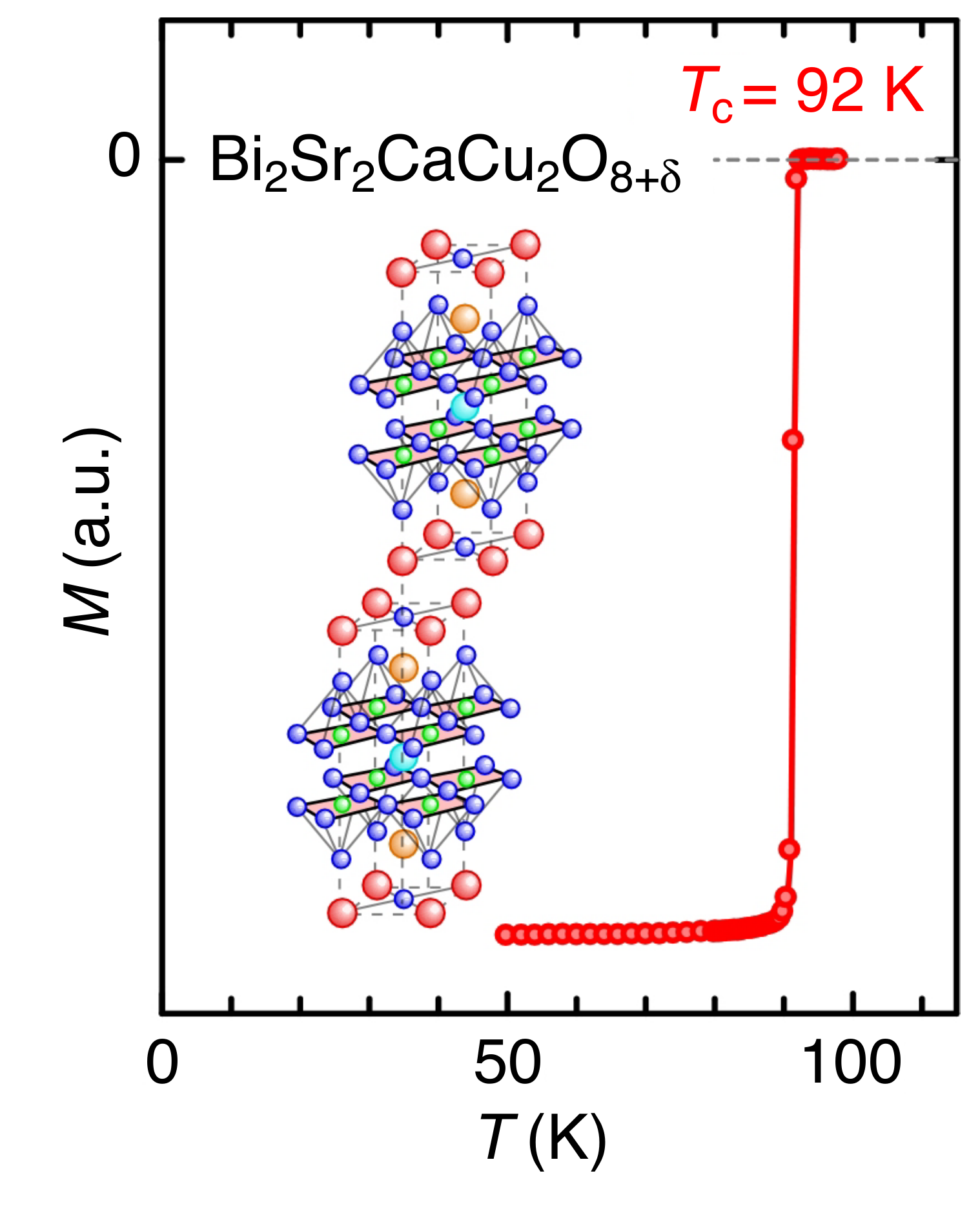}
\caption{Magnetic susceptibility for the single crystal of optimally doped Bi$_2$Sr$_2$CaCu$_2$O$_{8+\delta}$ (Bi2212)  with an onset $T_{\rm c}$ of 92K used for the ARPES measurements.  The crystal structure of Bi2212 is drawn in the inset. }
\label{SQUID}
\end{figure*}

\begin{figure*} 
\includegraphics[width=5.9in]{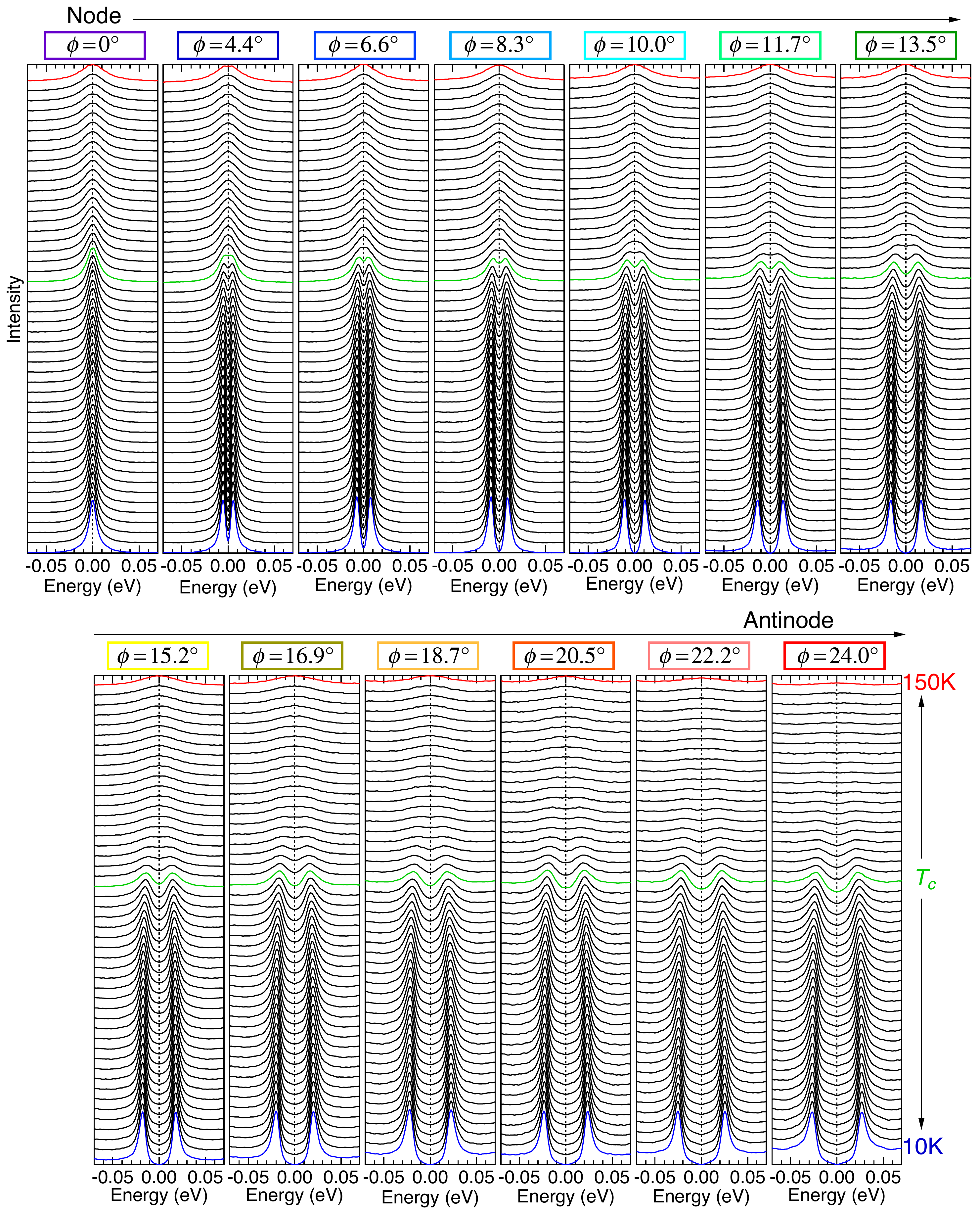}
\caption{Temperate evolution of ARPES spectra (symmetrized EDCs) at various $k_{\rm F}$s: the same data as in Fig.\ref{WithoutOffset}, but plotted with an offset. 
The top left panel plots the data at the node ($\phi  = 0^\circ$). Toward the bottom right, the measured $k_{\rm F}$ approaches the antinode ($\phi  = 45^\circ$). 
The blue, green, and red curves in each panel are the spectra at 10K, $T_c$ (=92K), and 150K, respectively. All the spectra at $T_c$, except for the nodal spectrum, have an energy gap as apparent from two peak structure, which signifies that the $d$-wave gap with a point node persists beyond the superconducting transition.}
\label{WithOffset}
\end{figure*}

\begin{figure*}  
\includegraphics[width=5.8in]{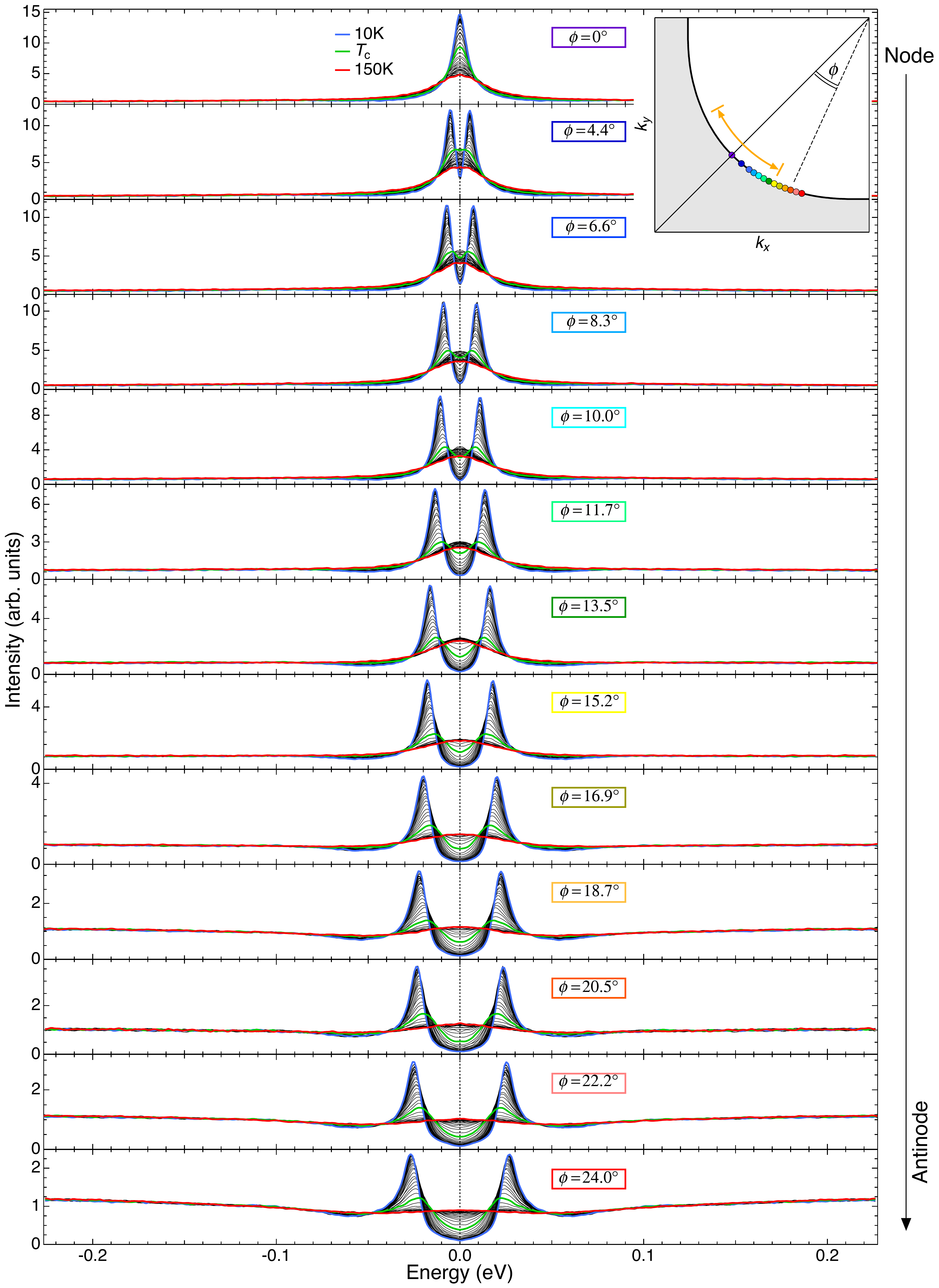}
\caption{Temperate evolution of ARPES spectra (symmetrized EDCs) at various $k_{\rm F}$s.  In the inset, the measured $k_{\rm F}$ points are 
marked on the Fermi surface with circles.
The  corresponding $\phi$ angles (defined in the inset)  are described in each panel.
The top panel plots the data at the node  ($\phi  = 0^\circ$). Toward the bottom, the measured $k_{\rm F}$ approaches the antinode ($\phi  = 45^\circ$). }
\label{WithoutOffset}
\end{figure*}

\begin{figure*} \label{}
\includegraphics[width=5.2 in]{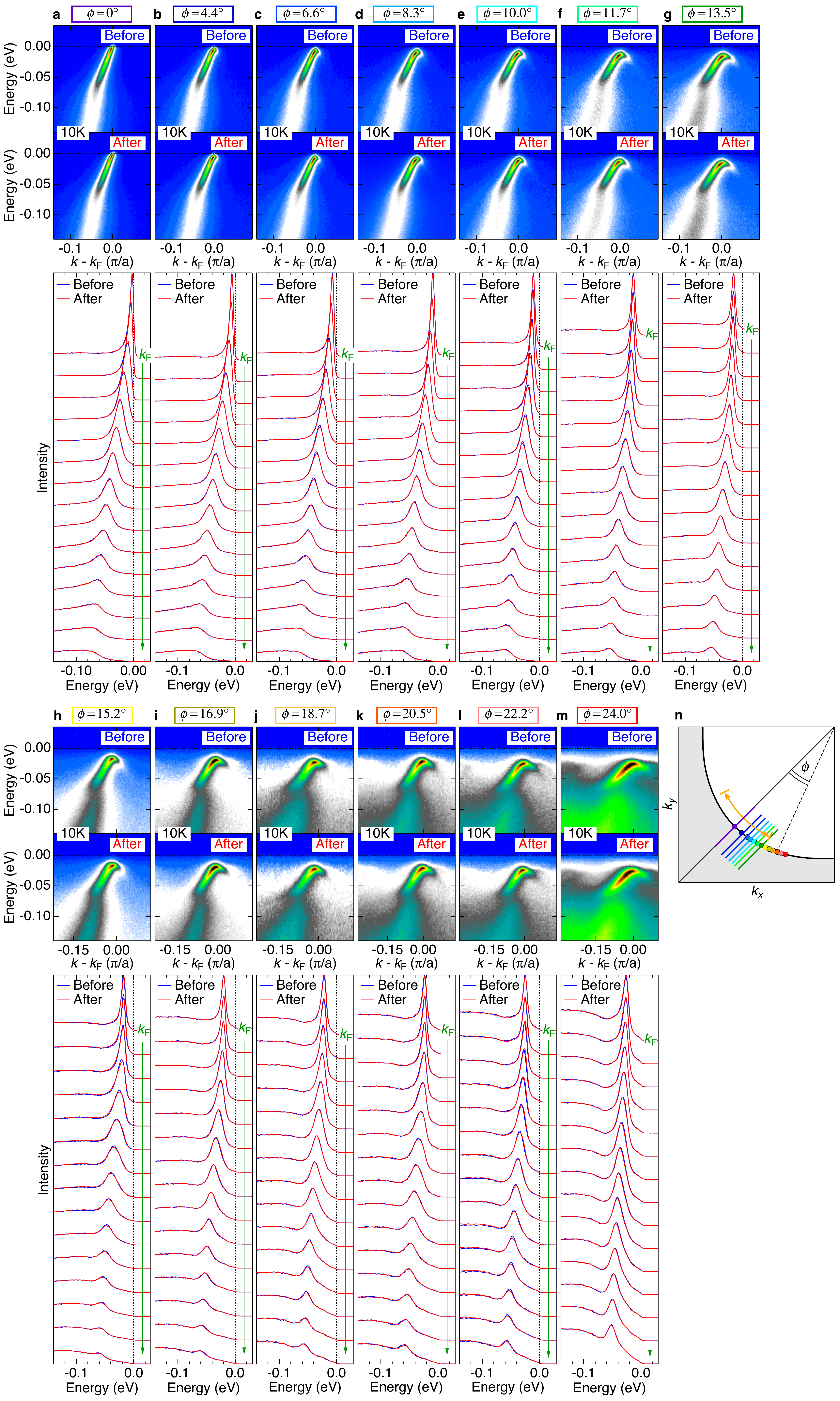}
\caption{
Aging check of the sample surface after temperature scan. The upper left panels in {\bf a} show the data along the node ($\phi  = 0^\circ$). Toward the bottom right in {\bf m}, the measured momentum approaches the antinode.  {\bf n}, Fermi surface with the measured momentum cuts and $k_{\rm F}$ points. 
} 
\label{AgingCheck}
\end{figure*}

\begin{figure*} 
\includegraphics[width=6.5in]{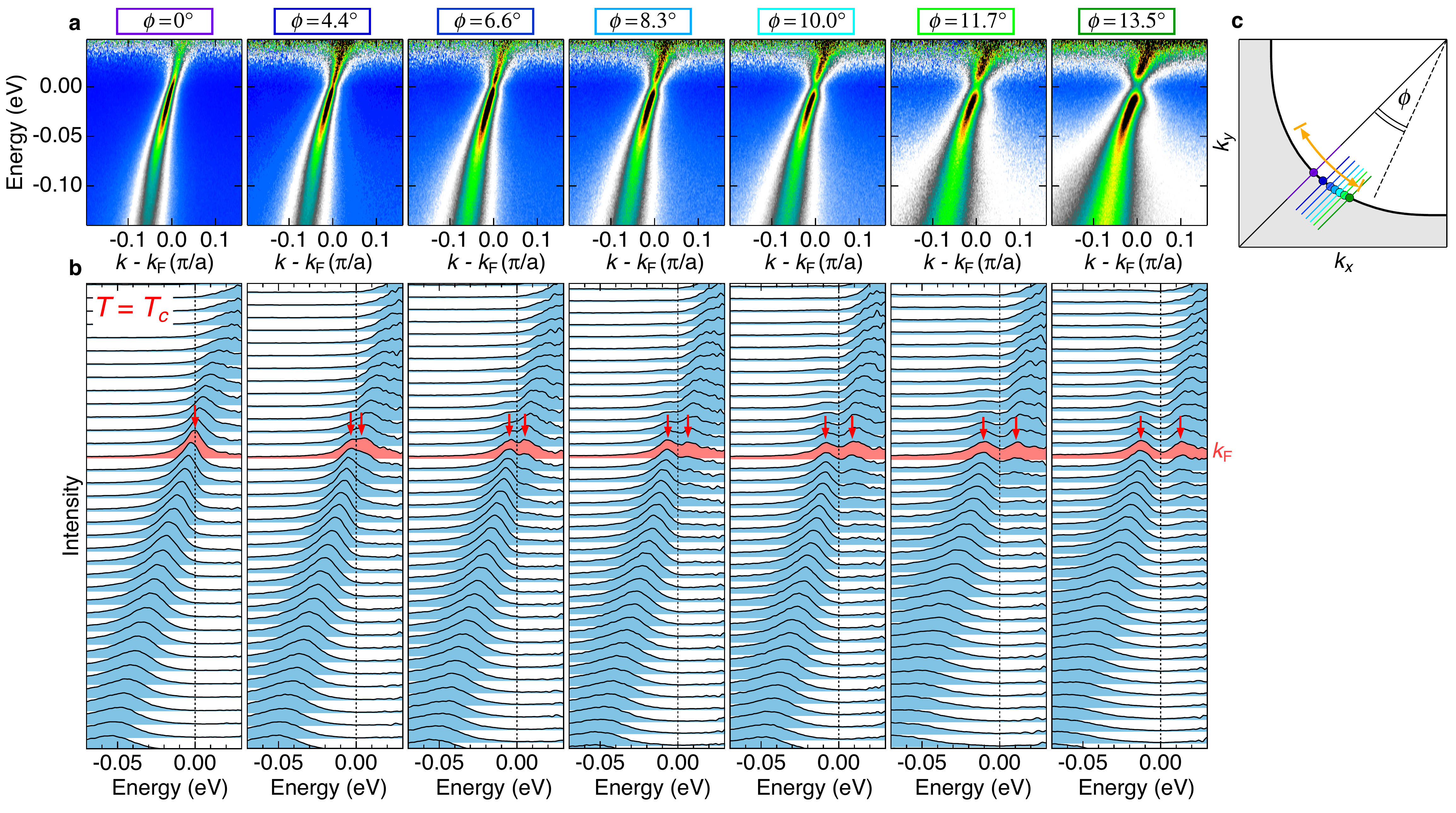}
\caption{Further evidence for the absence of the Fermi arc at $T_c$.  {\bf a},  Fermi-function divided band dispersions measured at $T_c$ along several momentum cuts (color lines in {\bf c}).  The same data  as Fig.2c in the main paper.
{\bf b},  The EDCs corresponding to each upper image in {\bf a}.  The spectra at $k_F$ points (color circles in {\bf c}) are painted with red. The peak positions are indicated with red arrows. 
{\bf c}, The Fermi surface. 
The orange arrow indicates the momentum region where the Fermi arc was previously claimed to emerge at $T_c$.
}
\label{Pointnode}
\end{figure*}

\begin{figure*} \label{}
\includegraphics[width=6.5in]{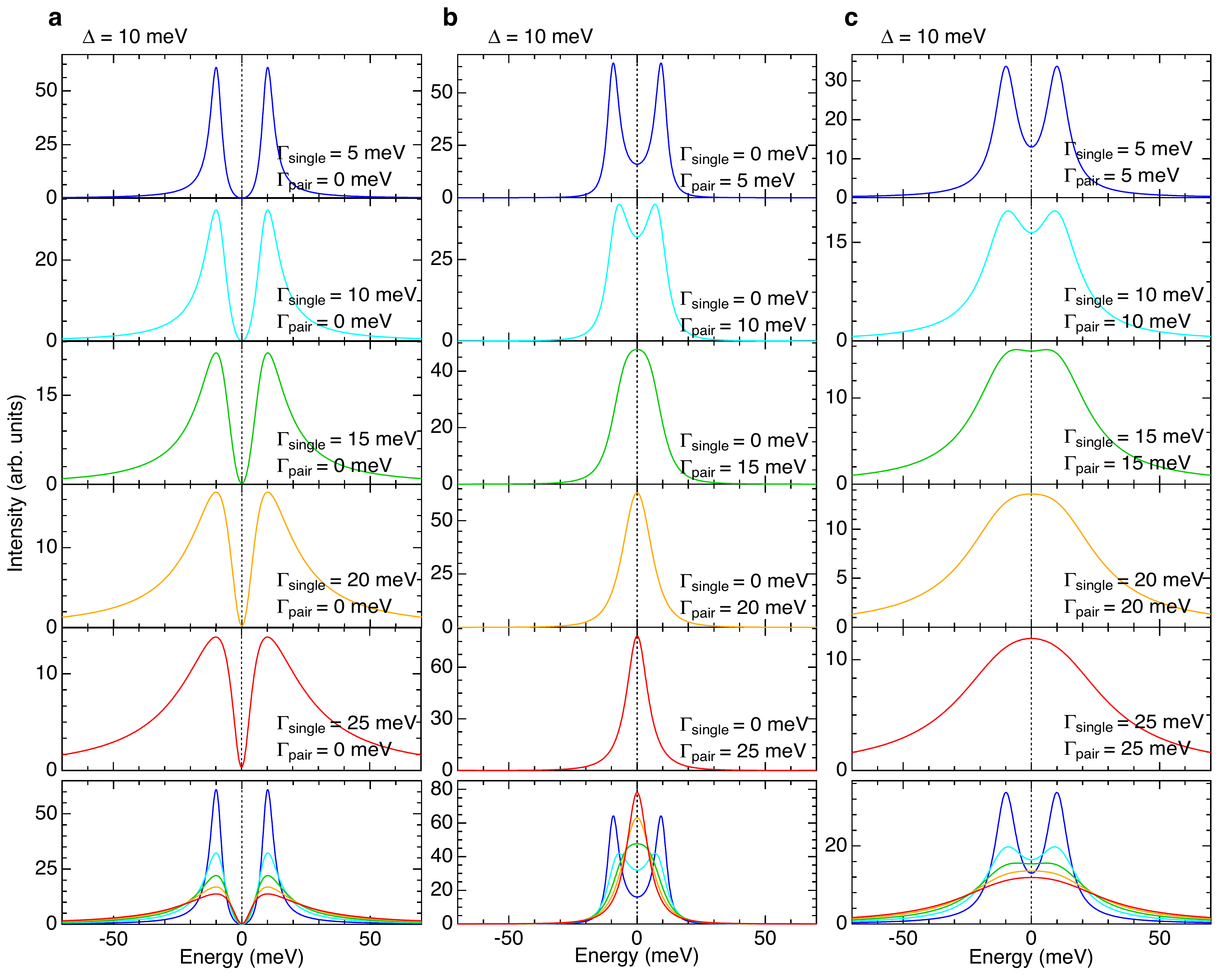}
\caption{Simulation of spectral function in Eq.(\ref{self_Norman}) for examining the effect of  
 two different scattering rates (${\Gamma _{{\text{single}}}}$ and ${\Gamma _{{\text{pair}}}}$) on the spectral shape.  
In all the curves presented, the gap magnitude is set to 10meV ($\Delta  \equiv 10$ meV). 
 {\bf a}, The ${\Gamma _{{\text{single}}}}$ dependence when ${\Gamma _{{\text{pair}}}}  \equiv 0$meV.
 {\bf b}, The ${\Gamma _{{\text{pair}}}}$ dependence when ${\Gamma _{{\text{single}}}} \equiv 0$meV.  
 {\bf c}, The spectral shape for various values of ${\Gamma _{{\text{single}}}}$ and ${\Gamma _{{\text{pair}}}}$, which  
are set to be equal. In each bottom panel of {\bf a}, {\bf b}, and {\bf c}, all spectra in the upper panels are superimposed. 
While all the curves plotted are convoluted with a gaussian that has the width of the experimental energy resolution ($\Delta \varepsilon  = 1.4$meV), the $\Delta \varepsilon$ value is so small that the difference in shape from original curves is negligible.} 
\label{Simulation}
\end{figure*}

\begin{figure*} \label{}
\includegraphics[width=6.5in]{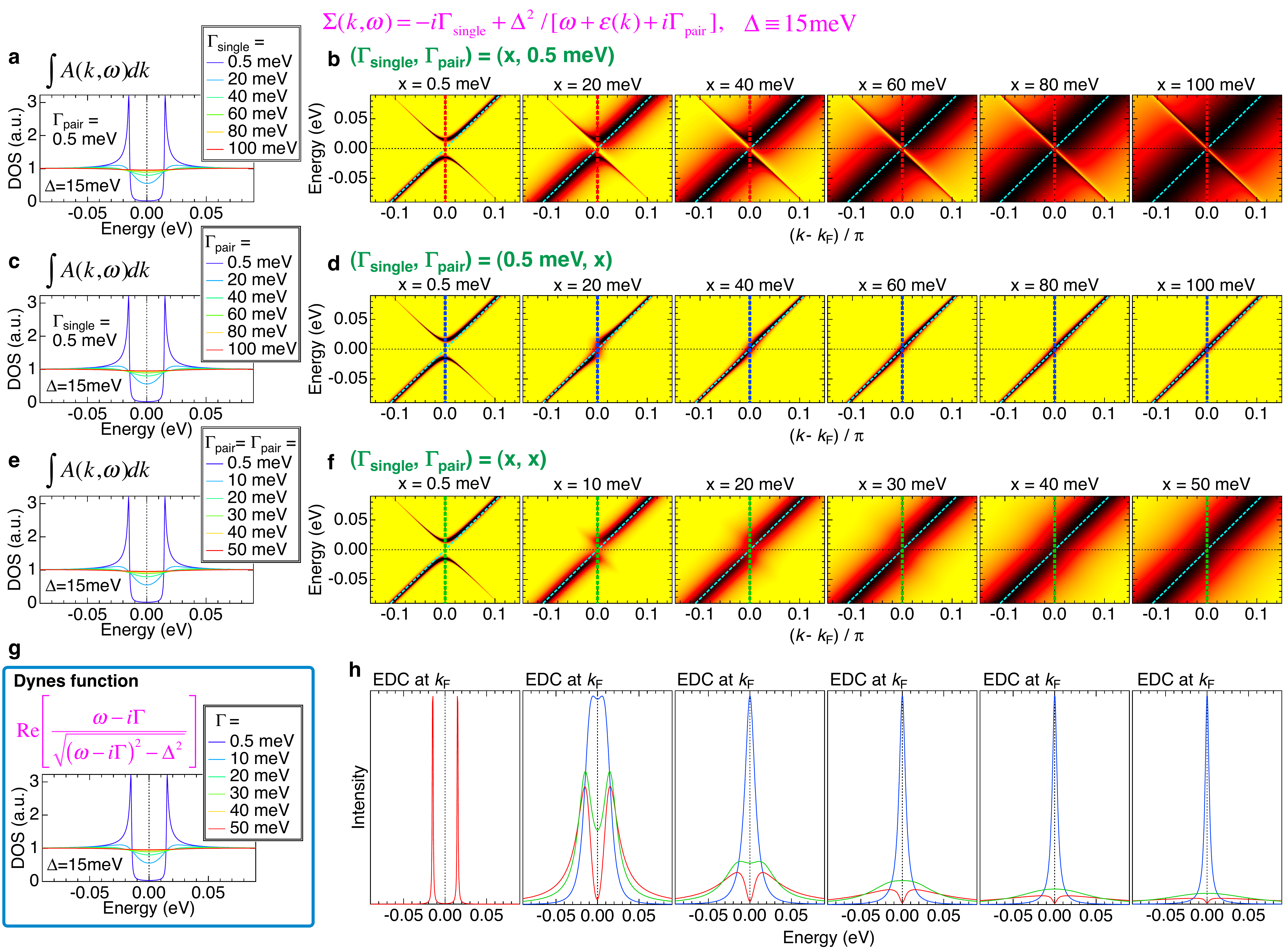}
\caption{Simulations for examining the effect of ${\Gamma _{{\text{single}}}}$ and ${\Gamma _{{\text{pair}}}}$ on the momentum-integrated spectra. The magnitude of energy gap $\Delta$ is fixed to 15 meV for all the cases.
{\bf a, c, e},  Momentum integrated spectra, ${I_{{\rm{DOS}}}}(\omega ) = \int {A(k,\omega )} dk$, 
 for the dispersion images in {\bf b, d, f}, respectively.
{\bf b, d, f}, Dispersion images for one-particle spectral function, $A(k,\omega )$, in Eq.(\ref{self_Norman}) with a linear $\varepsilon (k) $ (light blue dashed lines).
$\Gamma_{\rm single}$ and $\Gamma_{\rm pair}$ dependences are studied in ({\bf a, b}) and ({\bf c, d}), respectively. The case under the condition of ${\Gamma _{{\text{single}}}} = {\Gamma _{{\text{pair}}}}$ is simulated in ({\bf e, f}). {\bf g}, The spectra of Dynes function with several values of  $\Gamma$. 
{\bf h}, Spectra at $k_F$, $A(k_F,\omega )$, extracted from images of {\bf b, d, f}.
 Each color (red, blue, and green) in curves corresponds to that of thick dashed line added in the upper images of  {\bf b, d, f}.
}
\label{DOSsimulation}
\label{fig1}
\end{figure*}

\begin{figure*} \label{}
\includegraphics[width=6.5in]{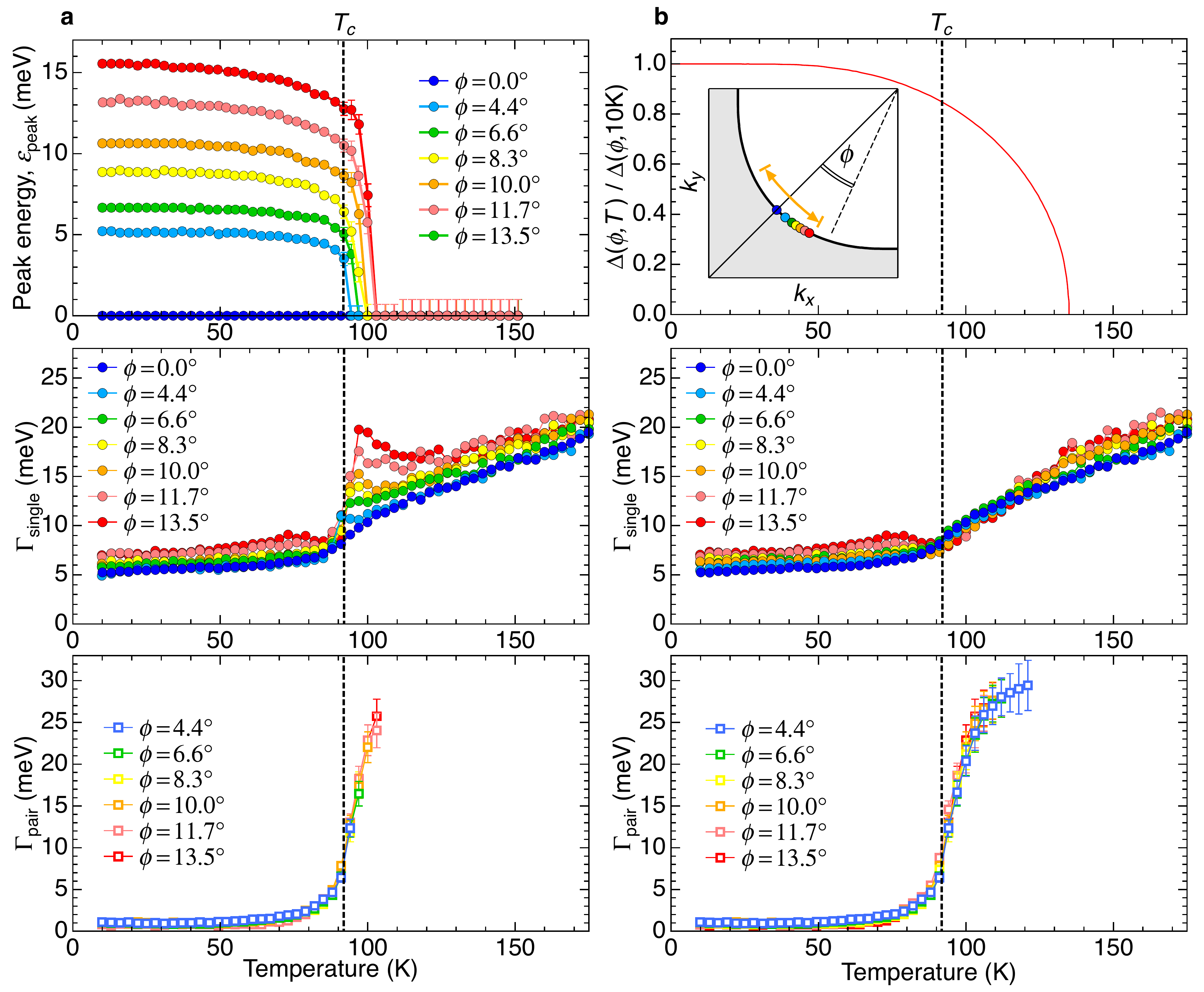}
\caption{Comparison of fitting results with Eq.(\ref{self_Norman})
under two different settings to the energy gap,  $\Delta $.  {\bf a}, The $\Delta $ is defined to be equal to the peak position of spectra [$\Delta (T) \equiv {\varepsilon _{{\text{peak}}}}(T)$], shown in the top panel. 
{\bf b}, The $\Delta$ is defined to have a BCS-type gap function with an onset at $T_{\rm pair}$=135K, shown in the top panel. 
The  middle and bottom panels plot 
the obtained fitting parameters,  ${\Gamma _{{\text{single}}}}$  and ${\Gamma _{{\text{pair}}}}$, at several $k_{\rm F}$ points in the momentum region, where the Fermi arc was previously claimed to emerge at $T_c$ (see circles and an arrow in the inset of {\bf b}). 
The values of $\Gamma_{\rm pair}$ at high temperatures are not plotted, since 
the spectral shape is not sensitive to the $\Gamma_{\rm pair}$ when $\Delta$ is small or zero, and thus it is impossible to precisely determine the value.
} 
\label{FittingResult}
\end{figure*}

\begin{figure*} \label{}
\includegraphics[width=5in]{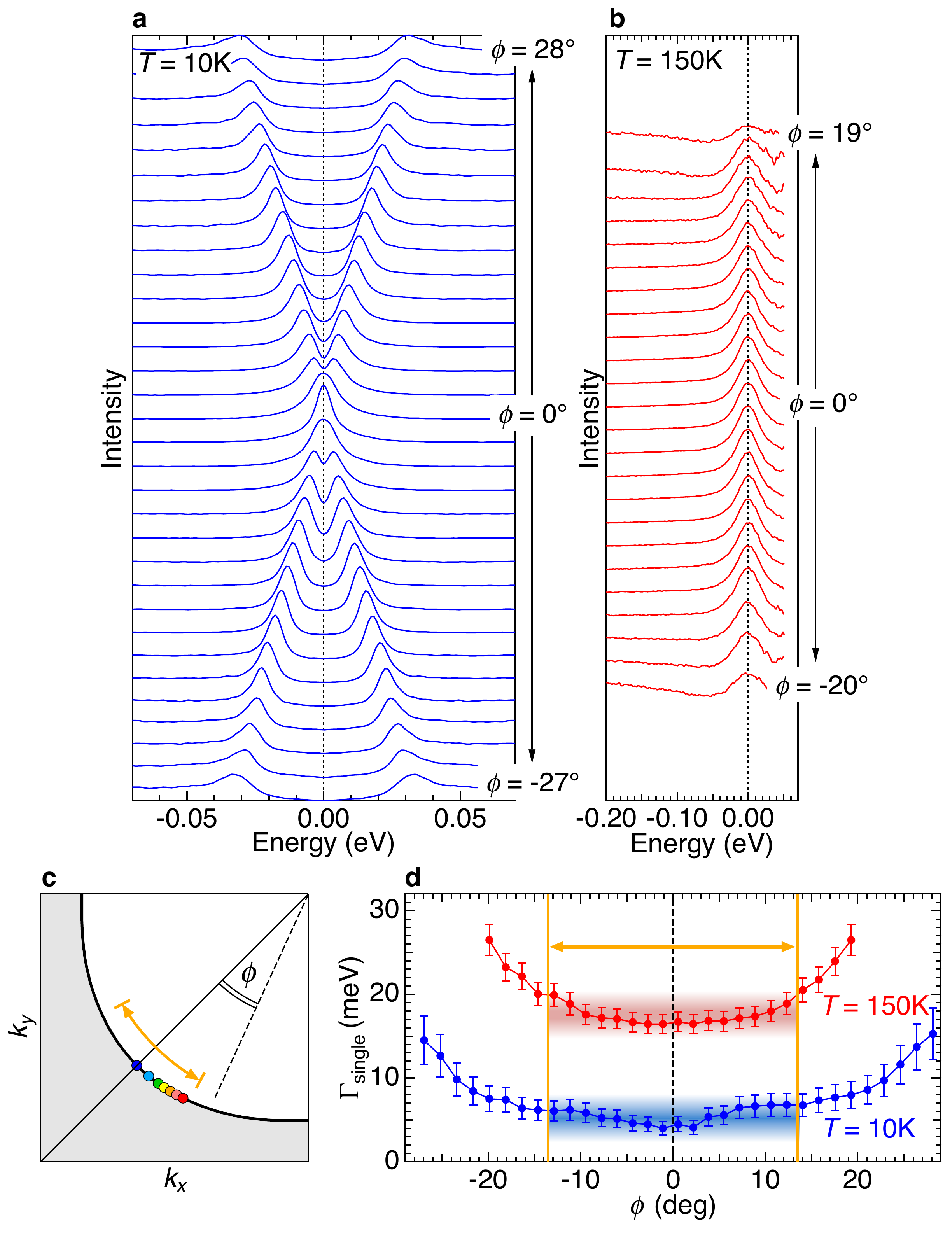}
\caption{Isotropic scattering mechanism around the node. 
{\bf a}, {\bf b},  ARPES spectra at various $k_{\rm F}$ points over a wide $\phi$ angle measured far below $T_c$ ($T$=10K) ({\bf a}) and   above $T_{\rm pair}$ ($T$=150K) ({\bf b}). In order to remove the effect of Fermi cut-off, the curves in {\bf a} are symmetrized about the Fermi level, and the ones in {\bf b} are divided by the Fermi function at 150K.  {\bf c}, The Fermi surface determined by tight binding fitting to the ARPES data. 
{\bf d}, The peak width of spectra (or ${\Gamma _{{\text{single}}}}$) in {\bf a} and {\bf b}. 
The orange arrows in {\bf c} and {\bf d} indicate the momentum region where the Fermi arc was previously claimed to emerge at $T_c$. The almost isotropic ${\Gamma _{{\text{single}}}}$ is obtained in this momentum region.  The momentum region studied in Fig.\ref{FittingResult} (marked by colored circles in {\bf c}) is located inside it. 
}
\label{IsotropicScattering}
\label{fig1}
\end{figure*}